\documentclass{article}
\usepackage{graphicx,amsmath,txfonts}
\usepackage{epstopdf, epsfig}
\usepackage[caption=false]{subfig}
\usepackage{natbib}
\usepackage{color}
\usepackage{authblk}

\begin{document}

\title{Signatures of quantum effects on radiation reaction in laser -- electron-beam collisions}

\author[1]{C.P. Ridgers}
\author[2]{T.G. Blackburn}
\author[1]{D. Del Sorbo}
\author[1]{L.E. Bradley}
\author[1]{C.D. Baird}
\author[3]{S.P.D. Mangles}
\author[4]{P. McKenna}
\author[2]{M. Marklund}
\author[1]{C. D. Murphy}
\author[5,6,7]{A.G.R. Thomas}
 
\affil[1]{York Plasma Institute, Department of Physics, University of York, Heslington, York, YO10 5DD, UK}
\affil[2]{Department of Physics, Chalmers University of Technology, SE-41296 Gothenburg, Sweden}
\affil[3]{The John Adams Institute for Accelerator Science, Blackett Laboratory, Imperial College London, South Kensington, London, SW7 2BZ, UK}
\affil[4]{Department of Physics SUPA, University of Strathclyde, Glasgow G4 0NG, United Kingdom}
\affil[5]{Center for Ultrafast Optical Science, University of Michigan, Ann Arbor, Michigan 48109-2099, USA}
\affil[6]{Cockcroft Institute, Daresbury Laboratory, Warrington WA4 4AD, United Kingdom}
\affil[7]{Department of Physics, Lancaster University, Lancaster LA1 4YB, United Kingdom}

\maketitle

\begin{abstract}
Two signatures of quantum effects on radiation reaction in the collision of a $\sim$ GeV electron-beam with a high-intensity ($>3\times10^{20}$\ Wcm$^{-2}$) laser-pulse have been considered.  We show that the decrease in the average energy of the electron-beam may be used to measure the Gaunt factor $g$ for synchrotron emission.  We derive an equation for the evolution of the variance in the energy of the electron-beam in the quantum regime, i.e. quantum efficiency parameter $\eta \nll 1$. We show that the evolution of the variance may be used as a direct measure of the quantum stochasticity of the radiation reaction and determine the parameter regime where this is observable.  For example, stochastic emission results in a 25\% increase in the standard deviation of the energy spectrum of a GeV electron beam, 1 fs after it collides with a laser pulse of intensity $10^{21}$\ Wcm$^{-2}$.  This effect should therefore be measurable using current high-intensity laser systems.
\end{abstract}

\section{Introduction}

Radiation reaction is the effective recoil force on an accelerating charged particle caused by the particle emitting electromagnetic radiation.  This effect will play an important role in laser-matter interactions at the intensities set to be reached by next generation high-intensity laser facilities ($\gtrsim10^{23}$\ Wcm$^{-2}$), where radiation reaction can lead to almost complete absorption of the laser-pulse: \cite{Bashinov_13} (using a classical theory) and \cite{Zhang_15} (including quantum corrections), have shown that radiation reaction gives an imaginary part in the dispersion relation for waves in a plasma. At intensities $\gtrsim 10^{23}$\ Wcm$^{-2}$, plasma electrons will become sufficiently energetic that in their individual rest frames the electric field $E_{RF}$ approaches the critical field for quantum electrodynamics $E_{\textnormal{crit}}=1.38\times10^{18}$\ Vm$^{-1}$ (\cite{Heisenberg_36}).  In this case, the emission of radiation by the electrons must be described in the framework of strong-field quantum-electrodynamics (QED), using the \cite{Furry_51} picture.  Specifically, when the quantum efficiency parameter $\eta=E_{RF}/E_{\textnormal{crit}}\gtrsim0.1$ the 
radiation reaction force becomes stochastic (\cite{Duclous_11}) and electron's dynamics are no longer well 
approximated by deterministic motion along a classical worldline (\cite{Shen_72}).

This quantum regime has been reached in experiments at CERN SPS in the interaction of $\sim 100$\ GeV electrons with the strong fields of atoms in a crystal lattice, as described by \cite{Andersen_12}, where the Gaunt factor for synchrotron emission was measured.  The analogous process of non-linear Compton scattering was studied experimentally at the Stanford Linear Accelerator (SLAC) in the interaction between an electron
beam of energy $E=46.6$\ GeV and a counter-propagating high-intensity ($10^{18}-10^{19}$\ Wcm$^{-2}$) laser-pulse, as reported by \cite{Bula_96} (positron generation was also observed in this experiment -- see \cite{Burke_97}).  In this experiment the laser intensity was too low to access the very non-linear regime of relevance to next generation laser-matter interactions, where $a_0\approx\sqrt{I\lambda^2/10^{18}\textnormal{\ Wcm}^{-2}\mu \mbox{m}^2}\gg1$ ($\lambda$ is the laser wavelength). This is now possible with current Petawatt laser systems, which can achieve focused intensities of $I>10^{21}$\ Wcm$^{-2}$.  In the interaction of an electron-beam with energy $\mathcal{E}$ with a counter-propagating laser-pulse of intensity $I$, $\eta$ can be estimated as $\eta \sim 0.1 (\mathcal{E}/500~\text{MeV}) \sqrt{I/10^{21}~\text{Wcm}^{-2}}$.  The quantum, non-linear regime of Compton scattering and the resultant radiation reaction can therefore be studied by accelerating the electrons to energies greater than 500MeV.  Laser wakefield acceleration (\cite{Tajima_79}) is a technique that can generate monoenergetic, well collimated and ultra-relativistic electron beams (\cite{Mangles_04},\cite{Geddes_04} \& \cite{Faure_04}). Recent experiments have now demonstrated energies approaching 5 GeV (\cite{Leemans_14}). Laser wakefield accelerators are ideal for studying electron beam collisions with the tightly focused lasers required for studies of nonlinear Compton scattering due to the inherent synchronicity of the generated electron beam and the laser which allows precise overlap in space and time. Therefore, all-optical equivalents of the SLAC experiment are possible using PW lasers (\cite{Sokolov_10_2,Thomas_12,Bulanov_12,Neitz_13,Blackburn_14,Vranic_14,Blackburn_15}). Non-linear Compton scattering at $a_0 \simeq 2$ (but not radiation reaction) was recently observed in such a setup by \cite{Sarri_14}.  Devising ways in which quantum effects on radiation reaction can be distinguished is therefore timely, as has been considered by \cite{DiPiazza_10,Neitz_13,Blackburn_14,Wang_15,Vranic_15,Harvey_17}.  

To simplify the treatment of quantum radiation reaction, we use the quasi-classical approach described by \cite{Baier_68}.  Here, we assume that the electromagnetic fields may be split into two types depending on their frequency scale.  Fields varying on the scale of the laser frequency are treated as classical background fields.  The photons emitted by the electrons on acceleration by these background fields, i.e. those responsible for the radiation reaction force, are treated in the framework of strong-field QED.  These photons are of much higher energy (typically $h\nu\gtrsim$\ MeV) than the laser photons ($h\nu\sim$\ eV).  Two further simplifying approximations are made (see \cite{Kirk_09}). By making the quasi-static approximation we assume that the formation length of the hard photons is much smaller than the scale over which the background fields vary and thus the background fields may thus be treated as constant over the space-time interval during which the emission occurs. This approximation is valid for $a_0\gg1$, which is the case in high-intensity laser matter interactions (\cite{DiPiazza_10} has shown that $a_0\gtrsim10$ is sufficient).  By making the weak-field approximation, we assume that the emission rate of photons depends entirely on $\eta$ and not the field invariants $\mathcal{F}=(E^2-c^2B^2)/E_{\textnormal{crit}}^2$ and $\mathcal{G}=c\mathbf{E}\cdot\mathbf{B}/E_{\textnormal{crit}}^2$.  This is valid if these invariants are much smaller than $\eta$.  For next-generation laser-matter interactions $E,cB\lesssim10^{-3}E_{\textnormal{crit}}$, so this approximation is also reasonable.  The weak-field approximation allows us to assume that the rate of photon emission (and the energy spectrum of the emitted photons) is well described by the well known rate in an equivalent set of constant fields as given in \cite{Ritus_85} (for constant crossed electric and magnetic fields) and \cite{Erber_66} (for a constant magnetic field).   The accuracy of this quasi-classical approach has recently been demonstrated  by comparison to full QED calculations for the electron energies and laser intensities considered here by \cite{Dinu_16}.

Using this quasi-classical model (making the quasi-static and weak-field approximations), it is possible to include the quantum radiation reaction force in a kinetic equation describing the evolution of the electron distribution, as given by \cite{Shen_72}, \cite{Elkina_11}, \cite{Sokolov_10_2}, \cite{Neitz_13} and \cite{Ridgers_14}.  Although this equation has been solved numerically using a Monte-Carlo algorithm (see \cite{Duclous_11,Elkina_11,Ridgers_14,Gonoskov_15}) it has not been solved analytically for even the simplest configuration of electromagnetic fields (for example a uniform, static magnetic field as in \cite{Shen_72}).  On the other hand, the electron equation of motion containing a classical model of radiation reaction, using the prescription of Landau \& Lifshitz (\cite{Landau_87} -- shown to be consistent with the classical limit of strong field QED by \cite{Krivitskii_91,Ilderton_13}), has been solved analytically in several cases for example: for electron motion in a rotating electric field (by \cite{Bell_08}) and a plane electromagnetic wave (by \cite{DiPiazza_08}).   A modified classical model, where the radiated power is reduced by the Gaunt factor,  has been used to derive the dispersion relation for an electromagnetic wave moving through a plasma where the electrons experience significant radiation reaction by \cite{Zhang_15} (and the equivalent classical result by \cite{Bashinov_13}).  The kinetic equation can be used to show that the modified classical model of radiation reaction is sufficient to describe the average energy loss of the electrons (\cite{Ridgers_14}). In addition, the kinetic equation can give insight into which observables can be used to measure various aspects of quantum radiation reaction. Here we show that the measurements of the average energy loss can be used to measure the Gaunt factor associated with the emission and that the evolution of the variance of the electron energy distribution can be used to measure the degree of stochasticity of the emission. To do the latter, we derive an equation of motion for the variance, which extends the results of \citet{Vranic_15} to arbitrary $\eta$.

 \section{Radiation reaction models}\label{sec:RR_Models}

In this section we describe the radiation reaction models considered here: (i) classical -- using the ultra-relativistic form of the Landau \& Lifshitz prescription; (ii) modified classical -- as the classical model but including a function describing the reduction in the power radiated due to quantum effects, the Gaunt factor $g$ (\cite{Baier_98}); (iii) stochastic -- a probabilistic treatment of the emission consistent with the approximations made in the quantum emission model described above and in more detail by \cite{Ridgers_14}.  The stochastic model is the most physical as it includes both the important quantum effects (the Gaunt factor and quantum stochasticity).  

Using the quasi-classical approach we may write the evolution of the electron distribution function, including the radiation reaction force, as

\begin{equation}
\label{Vlasov}
\frac{\partial f}{\partial t} + \mathbf{v}\cdot\frac{\partial f}{\partial{\mathbf{r}}} - e(\mathbf{E}+\mathbf{v}\times\mathbf{B})\cdot\frac{\partial f}{\partial{\mathbf{p}}} = \left(\frac{\partial f}{\partial t}\right)_{em}^X. \nonumber
\end{equation}

\noindent $f d^3\mathbf{x}d^3\mathbf{p}$ is the number of electrons at position $\mathbf{x}$ with momentum $\mathbf{p}$ (velocity $\mathbf{v}$).  $\mathbf{E}$ and $\mathbf{B}$ are the low frequency classical background electromagnetic fields. $(\partial f/\partial t)_{em}^X$ is an operator describing how recoil from photon emission affects the electron distribution function -- we will refer to this as the emission operator. The superscript $X$ denotes which of the classical ($cl)$, modified classical ($mod\ cl$) and stochastic ($st$) models is under consideration. 

Note that we are neglecting pair production by the emitted gamma-ray photons in the background electromagnetic fields.  This is reasonable in the moderately quantum regime described by \cite{DiPiazza_10}, i.e. where $\eta\sim 0.1$.

\subsection{Classical and modified classical emission operators}

If the radiating electron is ultra-relativistic with $\gamma \gg 1$, we may assume that all photons are emitted in the direction of the electron's instantaneous velocity (\cite{Duclous_11}). Using the Landau \& Lifshitz prescription for radiation reaction (in the ultra-relativistic limit -- \cite{Landau_87}) the classical and modified classical emission operators should describe radiation reaction forces of the form

\begin{equation}
\label{rad_reaction_force}
 \mathbf{F}_{cl} = -\frac{P_{cl}}{c} \hat{\mathbf{p}} \quad\quad \mathbf{F}_{mod \ cl} = -\frac{gP_{cl}}{c}\hat{\mathbf{p}}
\end{equation} 

\noindent respectively.  Here $g(\eta)$ is the Gaunt factor for synchrotron emission, i.e. a function that gives the reduction in the radiated power $P_{cl}$ due to quantum modifications to the synchrotron spectrum.  $P_{cl}$ is parameterised in terms of $\eta$ as

\begin{equation}
P_{cl} = \frac{2\alpha_f c}{3\lambdabar_c}m_ec^2\eta^2 \nonumber
\end{equation}

\noindent and $g(\eta)$ is defined as

\begin{equation}
g(\eta) = \frac{\int_0^{\eta/2}F(\eta,\chi)d\chi}{\int_0^{\infty}F_{cl}\left(\frac{4\chi}{3\eta^2}\right)d\chi} = \frac{3\sqrt{3}}{2\pi\eta^2}\int_0^{\eta/2}F(\eta,\chi)d\chi. \nonumber
\end{equation}

\noindent $F_{cl}$ and $F$ are the classical and quantum synchrotron spectra respectively.  For completeness their forms are given in appendix \ref{special_functions}.  An accurate fit to this function is $g(\eta)\approx [1 + 4.8(1+\eta)\ln(1+1.7\eta)+2.44\eta^2]^{-2/3}$ (\cite{Baier_98}). 
 
 The emission operators which yield radiation reaction forces as given in equation (\ref{rad_reaction_force}), as shown in section \ref{moments}, are

\begin{equation}
\label{deterministic_emission_operators}
\left(\frac{\partial f}{{\partial{t}}}\right)_{em}^{cl} = \frac{1}{p^2}\frac{\partial}{\partial p}\left( p^2 \frac{P_{cl}}{c}f \right) \quad\quad  \left(\frac{\partial f}{{\partial{t}}}\right)_{em}^{mod\ cl} = \frac{1}{p^2}\frac{\partial}{\partial p}\left( p^2 g \frac{P_{cl}}{c}f \right)
\end{equation}

\subsection{Stochastic emission operator}

The stochastic emission operator should consist of two terms: a term describing the movement of electrons out of a given region of phase space due to emission and a term describing electrons moving into the region under consideration by leaving regions of higher energy as they emit.  Assuming the electrons are ultra-relativistic and so photon emission is in the direction of propagation of the electron, we may formulate this as
 
\begin{equation}
\label{stochastic_emission_operator}
\left(\frac{\partial{}f}{\partial{}t}\right)_{em}^{st}=-\lambda_{\gamma}(\eta)f + \frac{b}{2m_ec}\int_{p}^{\infty}dp'\lambda_{\gamma}(\eta')\rho_{\chi}(\eta',\chi)\frac{p'^2}{p^2}f(\mathbf{p'}).
\end{equation}  

\noindent We define $\eta \equiv \gamma b$.  For $\gamma \gg 1$, we may take $b = |\mathbf{E}_{\perp} +\mathbf{v}\times\mathbf{B}|/E_s$.  $\chi=(h\nu b)/(2m_ec^2)$ is the quantum efficiency parameter for an emitted photon (with energy $h\nu$). The explicit form of the photon emission rate $\lambda_{\gamma}$ and the probability $\rho_{\chi}d\chi$ that an electron with energy parameterised by $\eta$ emits a gamma-ray photon with energy parameterised by $\chi$ are given in appendix \ref{special_functions}.

\section{Moment equations}
\label{moments}

The average over the distribution function $f$ of a momentum dependent quantity $\psi(\mathbf{p})$ is defined as 

\begin{equation}
\langle \psi(\mathbf{p}) \rangle \equiv \frac{1}{n_e}\int d^3\mathbf{p} \psi(\mathbf{p}) f(\mathbf{x},\mathbf{p},t). \nonumber
\end{equation}

\noindent where $n_e$ is the electron number density.

\subsection{The temporal evolution of $\langle\mathbf{p}\rangle$}

The equation for the evolution of the expectation value of the momentum of the electron population $\langle\mathbf{p}\rangle$ has been derived previously by \cite{Elkina_11}.  The equation for the evolution of the average energy $\langle\gamma\rangle$ of the population has been derived by \cite{Ridgers_14}:  

\begin{equation}
\label{first_moment_stochastic}
\left(\frac{d\langle\mathbf{p}\rangle}{dt}\right)_{st} = -\frac{\langle gP_{cl}\hat{\mathbf{p}}\rangle}{c}.
\end{equation}

\noindent In appendix \ref{moment_derivation} we show how this equation can be derived by taking the first moment of the stochastic emission operator in equation (\ref{stochastic_emission_operator}). 

Taking the first moment of the classical and modified classical emission operators given in equation (\ref{deterministic_emission_operators}), as detailed in appendix \ref{moment_derivation}, yields

\begin{equation}
\label{first_moment_deterministic}
\left(\frac{d\langle\mathbf{p}\rangle}{dt}\right)_{cl} = -\frac{\langle P_{cl}\hat{\mathbf{p}}\rangle}{c} \quad\quad  \left(\frac{d\langle\mathbf{p}\rangle}{dt}\right)_{mod \ cl} = -\frac{\langle gP_{cl}\hat{\mathbf{p}}\rangle}{c} 
\end{equation}

\subsection{The temporal evolution of $\sigma^2$}

Following the derivation in appendix \ref{moment_derivation} we can obtain the following equation for the evolution of the variance $\sigma^2$ in the Lorentz factor $\gamma$ of the electron distribution: 

\begin{equation}
\label{second_moment_stochastic}
\left(\frac{d\sigma^2}{dt}\right)_{st} = -2\frac{\langle\Delta\gamma g P_{cl}\rangle}{m_ec^2} + \frac{\langle S\rangle}{m_e^2c^4}.
\end{equation}  

\noindent $\sigma^2=\langle\gamma^2\rangle-\langle\gamma\rangle^2$ and $\Delta \gamma = \gamma - \langle\gamma\rangle$. The first term in equation (\ref{second_moment_stochastic}), which we label $T_{-}$, always acts to reduce the variance. It arises because higher energy electrons radiate more energy than those at lower energy. This term can be written $T_{-}=(2/m_ec^2)[\langle\Delta\gamma P_{cl}\rangle - \langle(1-g)\Delta\gamma P_{cl}\rangle]$, where the first term is purely classical and the second shows that quantum effects reduce the rate of decrease of the variance by reducing the power radiated below the classical prediction ($g\le 1$).  The second term in equation (\ref{second_moment_stochastic}) $T_{+}$ represents stochastic effects, is positive and so tends to increase the variance. The competition between these two terms determines whether the emission operator causes $\sigma^2(t)$ to increase or decrease.

The function $S(\eta)$ is given by

\begin{equation}
 \label{s_func}
S(\eta) = \frac{55\alpha_fc}{24\sqrt{3}\lambdabar_cb}m_e^2c^4\eta^4g_2(\eta). \nonumber
\end{equation}  

\noindent $g_2(\eta)$, which is analogous to $g(\eta)$, is defined as

\begin{equation}
g_2(\eta) = \frac{\int_0^{\eta/2} \chi F(\eta,\chi)d\chi}{\int_0^{\infty}\chi F_{cl}\left(\frac{4\chi}{3\eta^2}\right) d\chi} = \frac{144}{55\pi\eta^4}\int_0^{\eta/2}\chi F(\eta,\chi)d\chi. \nonumber
\end{equation}

\begin{figure}
\centering
\includegraphics[scale=0.45]{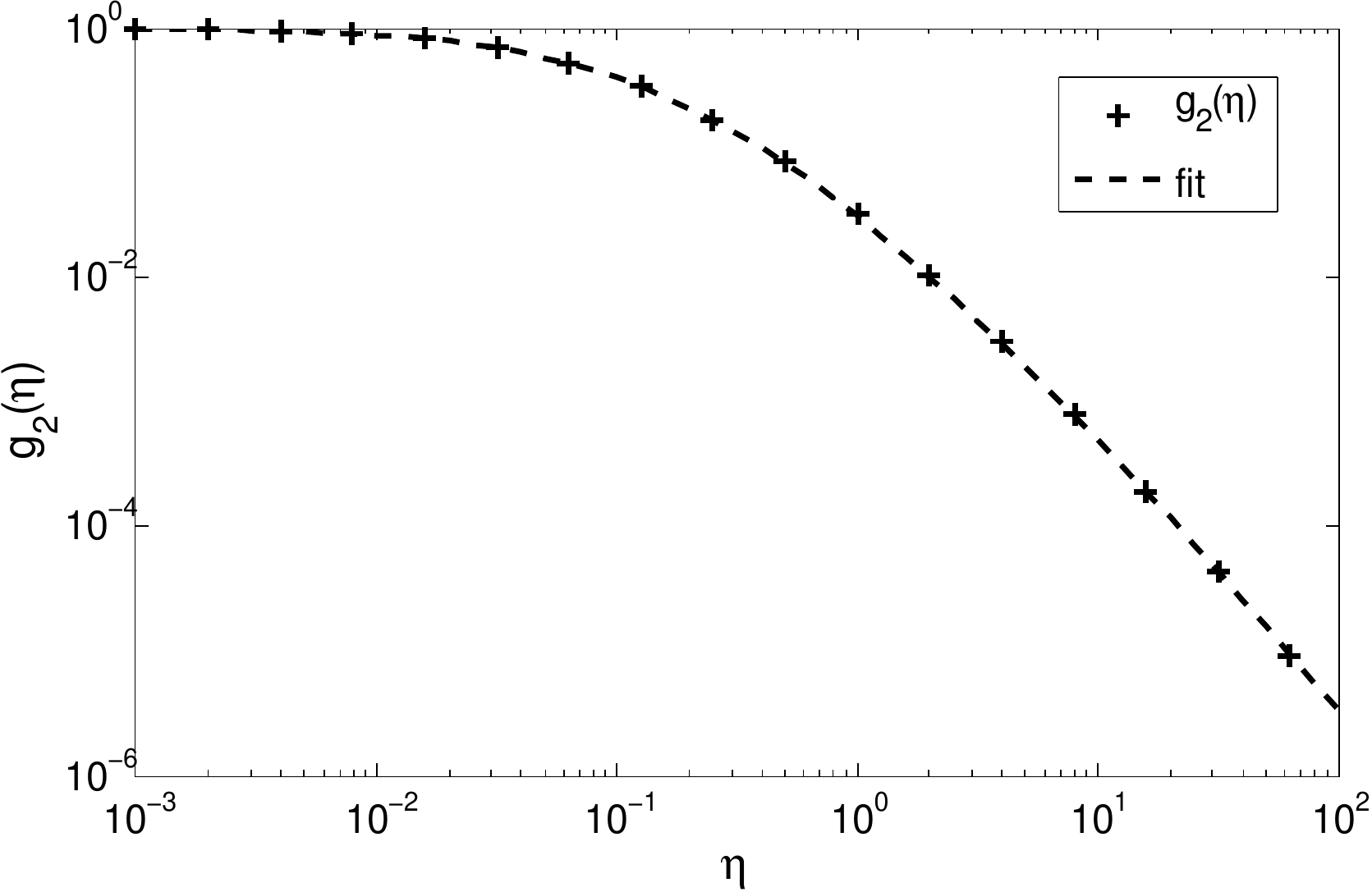}
\caption{\label{g_2_fig} $g_2(\eta)$ (solid line) and the fit used here (dashed line).}
\end{figure}

As for $g$, it is useful to find an accurate fit to $g_2$.  We find the following $g_2(\eta) \approx [1+(1+4.528\eta)\ln(1+12.29\eta)+4.632\eta^2]^{-7/6}$. This gives the correct limits for $\eta\ll1$ and $\eta\gg1$ ($g_2\approx1$ and $g_2\approx0.167\eta^{-7/3}$ respectively). $g_2$, as a function of $\eta$, along with the fit are shown in figure \ref{g_2_fig}.

We may also derive the corresponding expressions for $d\sigma^2/dt$ from the classical and modified classical emission operators in equation (\ref{deterministic_emission_operators}) (the derivation is given in appendix \ref{moment_derivation}). 

\begin{equation}
  \label{second_moment_deterministic}
\left(\frac{d\sigma^2}{dt}\right)_{cl} = -2\frac{\langle\Delta\gamma P_{cl}\rangle}{m_ec^2} \quad\quad \left(\frac{d\sigma^2}{dt}\right)_{mod \ cl} = -2\frac{\langle\Delta\gamma g P_{cl}\rangle}{m_ec^2}.
\end{equation}

We now consider the specific case where a high-energy electron beam with Gaussian energy distribution collides with a plane electromagnetic wave. In the limit where $\eta\ll1$ and the energy distribution is Gaussian with $\sigma \ll \langle\gamma\rangle$ (and assumed to be a Gaussian at all times), equation (\ref{second_moment_stochastic}) reduces to 

\begin{equation}
\label{V_low_eta}
\left(\frac{d\sigma^2}{dt}\right)_{st} \approx \frac{\alpha_fcb^2}{\lambdabar_c}\left(\frac{55b}{24\sqrt{3}}\langle\gamma\rangle^4-\frac{8}{3}\sigma^2\langle\gamma\rangle\right), \nonumber
\end{equation}

\noindent which reproduces equation 14 in \citet{Vranic_15}.

\section{Comparison to QED-PIC simulations}

To test the validity of the expression for the evolution of $\sigma^2$ given above we have simulated the interaction of an electron-beam with a counter-propagating circularly polarised plane-wave using the QED-PIC code {\small{EPOCH}} (\cite{Arber_15}). {\small{EPOCH}} includes the stochastic emission model using a Monte-Carlo algorithm (described in detail by \cite{Ridgers_14}).  For this work we have extended the code to include the classical and modified classical emission operators by directly solving equations (\ref{rad_reaction_force}) using first-order Eulerian integration.

The simulation parameters were as follows.  The laser pulse had peak intensity $10^{21}$\ Wcm$^{-2}$, wavelength 1~micron, and a half-Gaussian temporal profile (rise time 1~fs).  4000 grid cells were used to discretise a spatial domain extending from $-40$ microns to $40$ microns. $10^5$ macroparticles were used to represent an electron bunch consisting of $10^9$ electrons.  The electron bunch had a Gaussian spatial profile, centred on 39.7 microns, with a FWHM of 0.17 microns and had initial distribution $f(\mathbf{x},\mathbf{p},t=0)=[n_e(\mathbf{x})/(\sqrt{2\pi}\sigma)]\delta(p_y)\delta(p_z)\exp[-(p_x+\gamma_0m_ec)^2/(2\sigma^2)]$ where $\mathbf{p}=(p_x,p_y,p_z)$ is the momentum coordinate in phase space and $n_e$ the number density of electrons in the beam.  $\gamma_0$ was the initial average energy of the bunch. 

\begin{figure}
\centering
\includegraphics[scale=0.5]{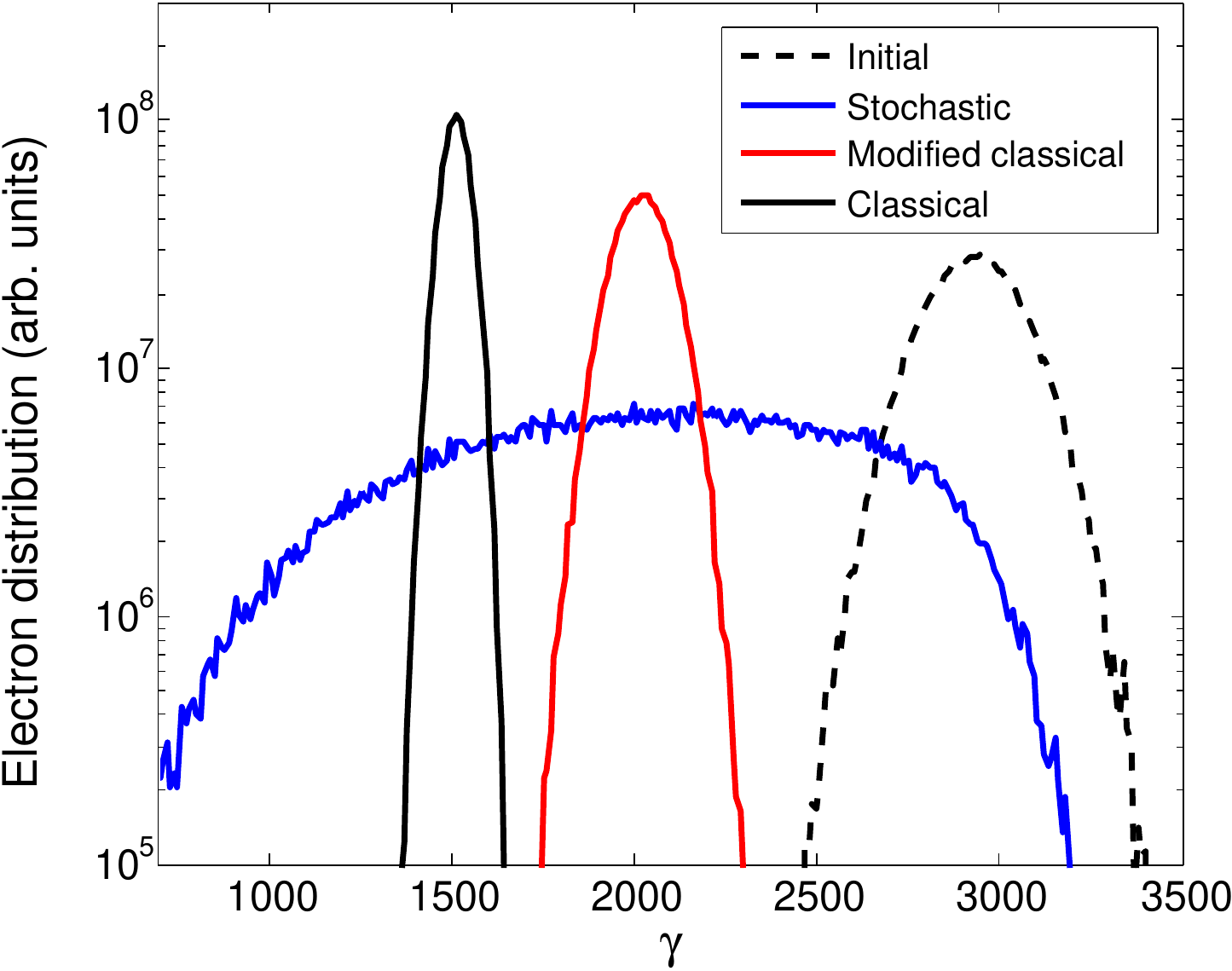}
\caption{\label{energy_dist_snapshot} Electron energy distribution, $10.5$\ fs after collision of the electron bunch with the laser-pulse, compared to initial distribution using the stochastic, modified classical and classical emission operators.}
\end{figure}

\begin{figure}
\centering
\includegraphics[scale=0.7]{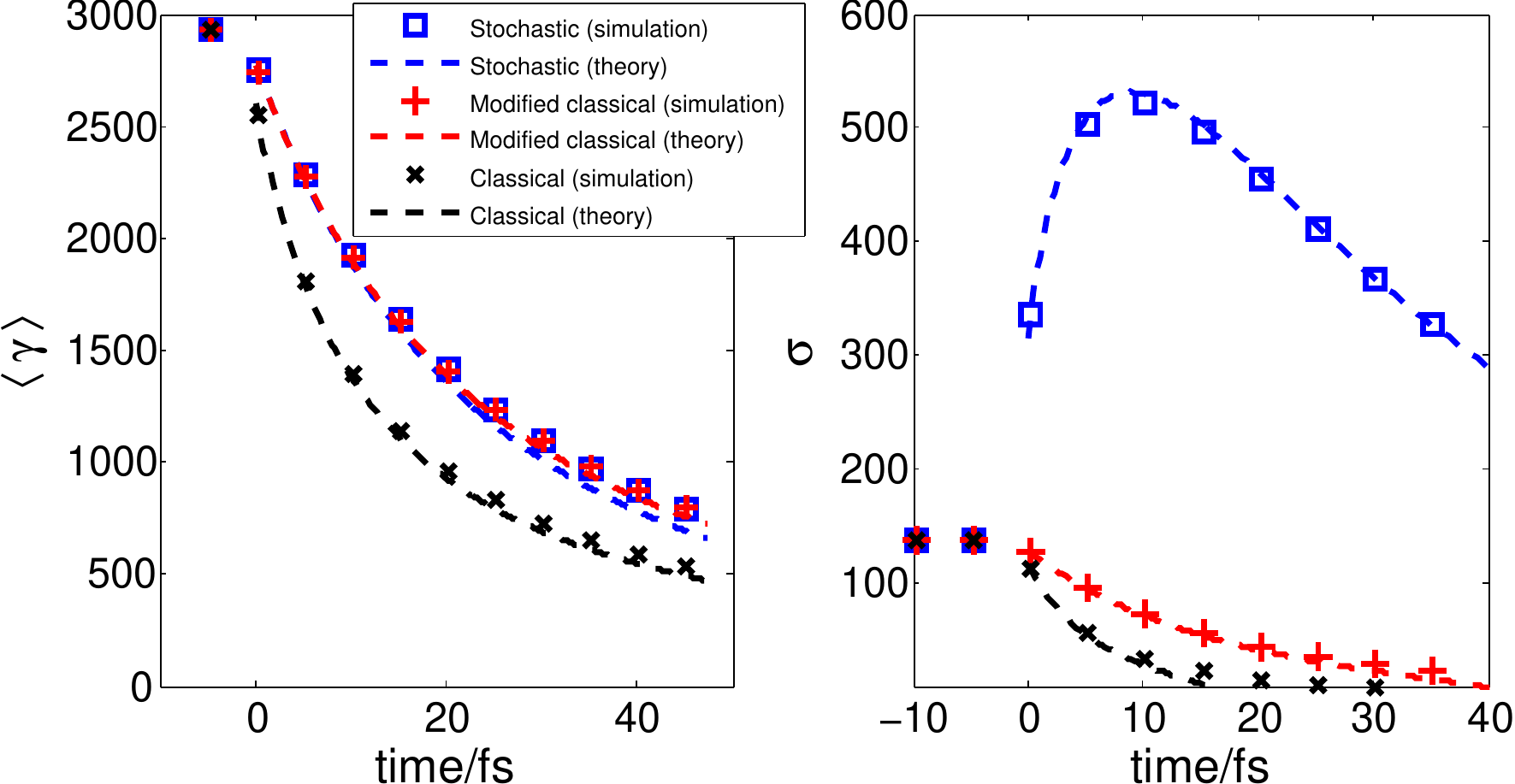}
\caption{\label{mean_var_time} Left: mean Lorentz factor versus time using the various emission models from simulation and as predicted by equations (\ref{first_moment_stochastic}) and (\ref{first_moment_deterministic}). Right: standard deviation in Lorentz factor versus time from simulation and as predicted by equations (\ref{second_moment_stochastic}) and (\ref{second_moment_deterministic}).}
\end{figure}

Figure \ref{energy_dist_snapshot} shows a comparison of the spatially integrated electron energy distribution using classical, modified classical and stochastic emission operators with the initial spectrum $t=10.5$\ fs after the collision.  We see that the modified classical and classical emission operators both give a decrease in the variance of the electron distribution whereas the stochastic emission operator gives an increase in the variance.  Figure \ref{mean_var_time} shows the temporal evolution of the mean Lorentz factor $\langle\gamma\rangle$ and the standard deviation of the Lorentz factor $\sigma$.  The QED-PIC simulations demonstrate the validity of equations (\ref{first_moment_stochastic}), (\ref{first_moment_deterministic}), (\ref{second_moment_stochastic}) \& (\ref{second_moment_deterministic}).

We saw earlier in equation (\ref{second_moment_stochastic}) that the evolution of the variance is governed by the competition between $T_{-}$ and $T_{+}$. To characterise which of these terms is dominant (in a similar way to \cite{Vranic_15}), and thereby how stochastic quantum effects (prevalent when $T_{+}$ dominates), may be measured in a colliding beams experiment, we derive an analytical expression their ratio $\xi$:  

\begin{equation}
\xi = \frac{T_{+}}{T_{-}} \quad\quad T_{+} = \frac{\langle S\rangle}{m_e^2c^4} \quad\quad T_{-} = 2\frac{\langle \Delta\gamma g P_{cl}\rangle}{m_ec^2}. \nonumber
\end{equation}

Considering an electron bunch whose initial distribution is $f(\mathbf{x},p_x,t=0)= n_e(\mathbf{x})/(2W\gamma_0m_ec)\delta(p_y)\delta(p_z)$ for $\gamma_0m_ec(1-W)<|p_x|<\gamma_0m_ec(1+W)$ and assuming $g=g_2=1$, we obtain (as outlined in appendix \ref{xi_deriv})

\begin{equation}
\label{T_Q_over_T_C}
\xi \approx (3.0 + 1.5W^{-2} + 0.3 W^2) \eta_0
\end{equation}

\noindent  where $\xi$ is the ratio $T_{+}/T_{-}$ when the electron bunch first collides with the laser pulse (i.e. before the distribution $f$ has evolved under the action of radiation reaction) and $\eta_0=\gamma_0b$.  As the variance increases and the expectation value of the $\gamma$ decreases we expect $T_{-}$ to eventually become dominant and so we would expect the variance to peak and then decrease after some time.  This behaviour is clearly seen in the results from the simulation using the stochastic emission operator shown in figure \ref{mean_var_time}.   Therefore, we define $T_{+}$ as being important for $\xi>2$ initially in order to compensate for the increased importance of $T_{-}$ at later times. In the case where the width of the electron distribution is equal to the mean, $W=0.5$, equation (\ref{T_Q_over_T_C}) shows that $\eta_0>0.2$ is required for $\xi>2$. For a narrow electron distribution, $W\ll1$, $\eta_0>1.3W^2$ is required and so $T_{+}$ can be important at lower $\eta_0$.

\begin{figure}
\centering
\includegraphics[scale=0.7]{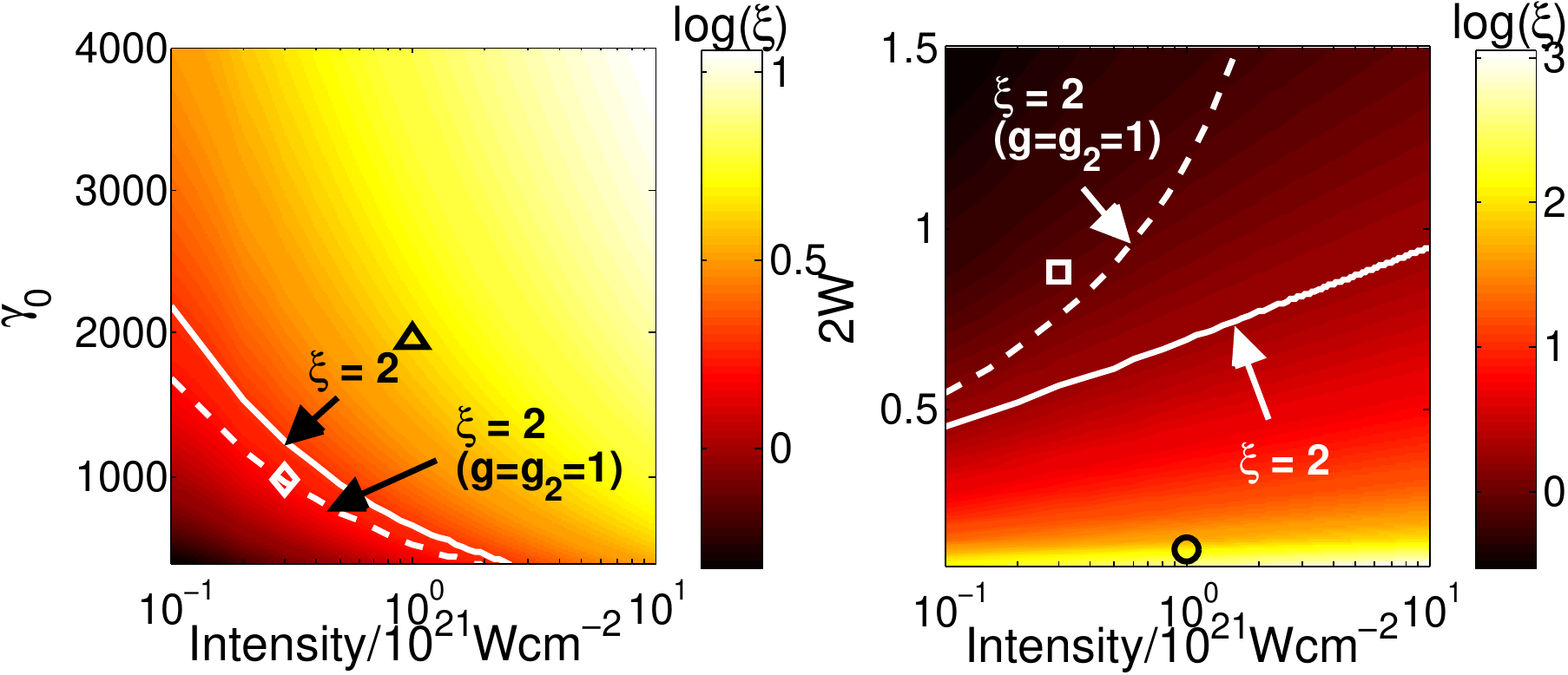}
\caption{\label{xi_gamma_I} $\xi$ as a function of: laser intensity and average Lorentz factor of the electron bunch (left); laser intensity and width of the electron energy distribution (right).  The solid white lines show $\xi=2$ and the dashed white lines show the prediction of where $\xi=2$ from equation (\ref{T_Q_over_T_C}).}
\end{figure}

From equation (\ref{T_Q_over_T_C}) we see that $\xi$ depends on three variables: the average Lorentz factor of the electron bunch $\gamma_0$; the width of the electron energy distribution $W$ and the laser intensity $I$ (which determines $b$).  Figure \ref{xi_gamma_I} shows $\xi$ (including $g$ \& $g_2$) as a function of $I$ \& $\gamma_0$ (for $W=0.2$) and $W$ \& $I$ (for $\gamma_0m_ec^2=1.5$\ GeV).  The prediction of $\xi=2$ from equation (\ref{T_Q_over_T_C}), i.e. making the assumption $g=g_2=1$, is shown to be reasonably accurate for $I\lesssim10^{21}$\ Wcm$^{-2}$.

To investigate whether the expression for $\xi$ in equation (\ref{T_Q_over_T_C}) predicts whether $T_{+}$ or $T_{-}$ dominates the evolution of the variance we performed further {\small{EPOCH}} simulations of the interaction of an electron-beam (again with initial distribution $f(\mathbf{x},\mathbf{p},t=0)=[n_e/(\sqrt{2\pi}\sigma)]\delta(p_y)\delta(p_z)\exp[-(p_x+\gamma_0m_ec)^2/(2\sigma)^2]$) and a counter-propagating plane-wave of intensity $I$.  The following parameters were chosen:
\ \\

\begin{center}
\begin{tabular}{ c | c | c | c | c } 
 Simulation & $I/10^{21}$Wcm$^{-2}$ & \ $\gamma_0m_ec^2$/GeV \  & FWHM/GeV & Symbol\\
 1 & 1.0 & 1.0 & 0.81  & $\triangle$ \\ 
 2 & 0.3 & 0.5 & 0.21 & $\Diamond$ \\ 
 3 & 1.0 & 1.5 & 0.17 & o \\ 
 4 & 0.3 & 1.5 & 1.3 & $\square$ \\ 
\end{tabular}
\end{center}

\ \\

\begin{figure}
\centering
\includegraphics[scale=0.6]{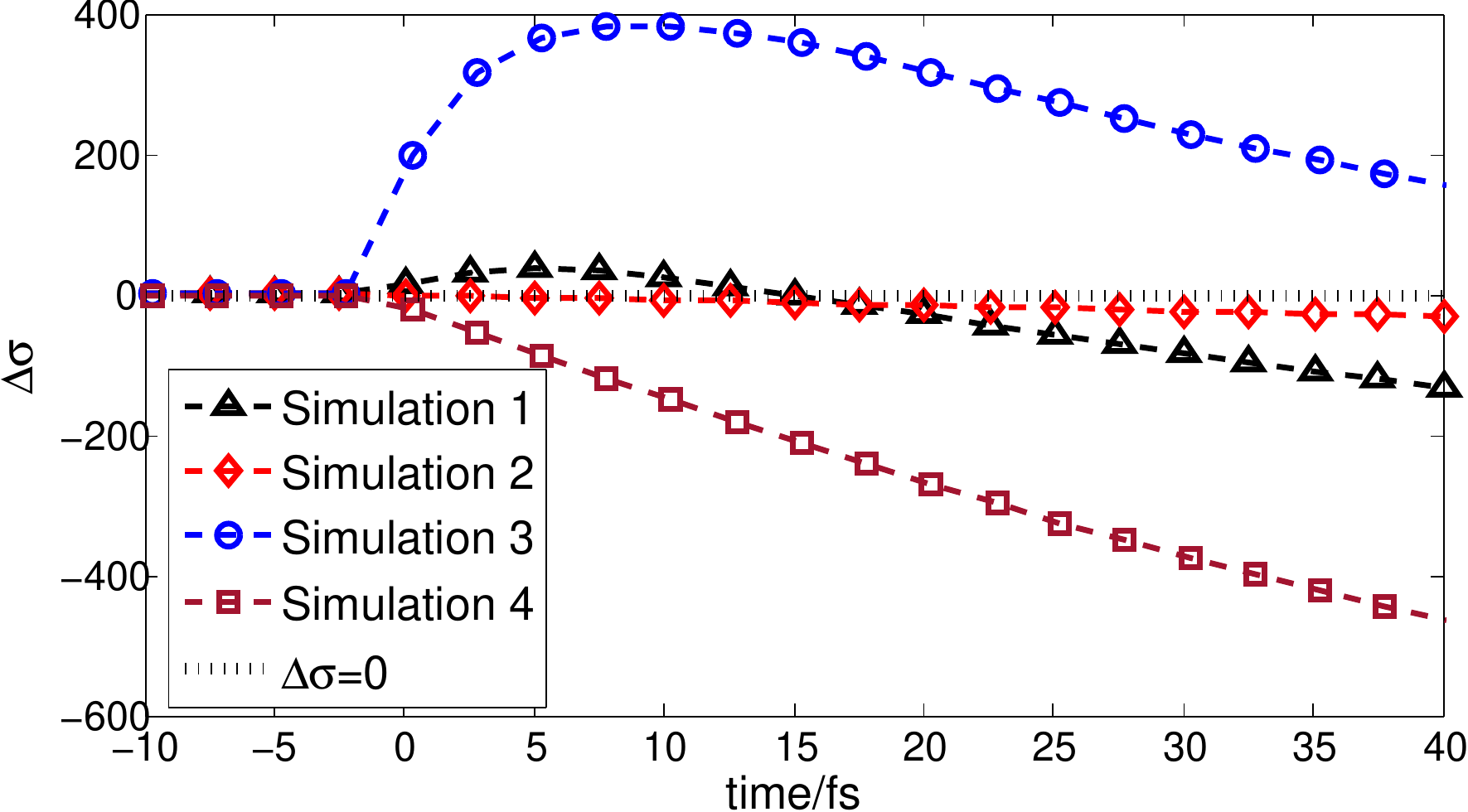}
\caption{\label{var_evolution} Temporal evolution of the change in standard deviation in the electron energy distribution in simulations 1--4.}
\end{figure}

We have shown where these simulations lie in the parameter space shown in figure \ref{xi_gamma_I} according to the symbols given in the table and assuming $W=\sqrt{2}\sigma$.  The time evolution of the change in the standard deviation of the electron energy distribution in these simulations is shown in figure \ref{var_evolution}.  We see that only those simulations where equation (\ref{T_Q_over_T_C}) predicts that $T_{+}$ is dominant show an increase in the variance.


\section{Discussion}

The results of this investigation can be summarised as follows: 

\begin{enumerate}
\item{$\langle\mathbf{p}\rangle$ evolves in the same way for the stochastic and modified classical emission operators and differently for the classical emission operator.}
\item{$\sigma^2$ evolves differently for all operators.  In particular, the stochastic emission operator can result in an increase in $\sigma^2$ whereas the classical and modified classical operators can only cause a decrease in $\sigma^2$ (as seen by \cite{Vranic_15} for $\eta\ll1$).}
\end{enumerate}

Result (i) requires further explanation.  Although we have shown that $(d\langle\mathbf{p}\rangle/dt)_{st}$ and $(d\langle\mathbf{p}\rangle/dt)_{mod\ cl}$ evolve according to the same equation, it does not necessarily follow that the expectation values themselves are the same for these two emission models (as noted by \cite{Elkina_11}).  We have previously shown in \cite{Ridgers_14} that, in fact, the expectation values of the energy using these two models do agree to a high degree of accuracy and this was shown again for the parameters considered here in figure \ref{mean_var_time}.  We would expect this in the classical limit where $\eta\ll1$.  In this case $T_{-}$ in equation (\ref{second_moment_stochastic}) dominates (from equation (\ref{T_Q_over_T_C}) we see that $\xi\propto\eta_0$) and rapidly reduces the variance of the electron bunch; the electron distribution in both the modified classical and stochastic models approaches a delta-function $\delta(\mathbf{p}-\langle\mathbf{p}\rangle)$.  The time evolution of $\langle\mathbf{p}\rangle$ depends on $\langle P_{cl}\rangle$ ($g \approx 1$ in the classical limit) which is equal to $(\langle\eta\rangle)P_{cl}(\langle\eta\rangle)$ for both the stochastic and modified classical models when $f$ is narrow in momentum-space.  However, in the simulation whose results are shown in figure \ref{mean_var_time} $\eta>0.1$.  From figure \ref{energy_dist_snapshot} we see that in this case the electron energy distribution is very different when the stochastic emission operator is used compared to when the modified classical emission operator is used.  Despite this the evolution of $\langle\mathbf{p}\rangle$ is the same due to the functional form of $gP_{cl}$.  When $\eta\gg1$, $gP_{cl}\propto\eta^{2/3}$.  This almost linear dependence on $\eta$ means that the difference in the evolution of $\langle\mathbf{p}\rangle$ between the models should be small.  Finally we note that, as shown in figure \ref{mean_var_time},  $\langle\mathbf{p}\rangle$ predicted by the classical emission model differs markedly from that predicted by the modified classical and stochastic models due to the neglect of the Gaunt factor $g$ in the classical model.  

$d\sigma^2/dt$ is always negative for both the classical and modified classical emission operators.  Physically, this is because electrons at higher energy radiate more energy than those at lower energy, causing a decrease in the width of the energy distribution.  The classical operator predicts a more rapid decrease than the modified classical operator due to the assumption that $g=1$ and the consequent overestimate of the scaling of the power radiated by the electrons with increasing $\eta$. For the stochastic emission operator $d\sigma^2/dt$ can be either positive or negative and so $\sigma^2$ can increase or decrease.  The evolution of $\sigma^2$ is determined by the balance between $T_{+}$ (which causes $\sigma^2$ to increase due the probabilistic nature of the emission) and $T_{-}$ (which, as just described, causes $\sigma^2$ to decrease as higher energy electrons radiate more energy).  We have shown (as did \cite{Vranic_15}) that which of these terms dominates depends on the width of the energy distribution and $\eta$. For large width $T_{-}$ increases in importance as it depends on $\Delta\gamma=\gamma-\langle\gamma\rangle$.  For high $\eta$ $T_{+}$ becomes more important due to its scaling with $\eta^4$ compared to at most $\eta^3$ for $T_{-}$ (assuming $\Delta\gamma\sim\gamma$).  In equation (\ref{T_Q_over_T_C}) we have provided a formula for the determination of which term is dominant.  
 
The first of these results, i.e. that the evolution of the expectation value is the same for the modified classical and stochastic (but not classical) models, is useful in two ways.  Firstly it shows that measuring the expectation value of an electron bunch after interaction with a high-intensity laser-pulse can give information about one quantum effect: the reduction of the total power emitted as expressed by $g$.  It cannot, however, give information about the probabilistic nature of the emission. Secondly, this result suggests that the modified classical model of radiation reaction is sufficient for the calculation of laser absorption in high-intensity laser-plasma interactions \cite{Brady_12,Zhang_15}.  Laser absorption in this context depends on the average energy loss by the electrons (and positrons) in the plasma due to radiation reaction.  The second result, i.e. the evolution of the variance differs between the models, can be used to measure the stochasticity of the radiation reaction.  An increase in the variance of the energy distribution of electrons must be due to the probabilistic nature of the emission.  As further work we propose a comparison of QED-PIC simulations of laser absorption in laser-plasma interactions using the different emission models and an investigation of the use of the variance to observe stochasticity in 3D simulations of the interaction of a focusing laser-pulse with a counter propagating electron bunch produced by laser wakefield acceleration (with a realistic energy spectrum).  

\section{Conclusions}

We have derived equations for the evolution of the expectation value of the momentum and variance in the energy of an electron population subject to three different radiation reaction models.  We have considered classical and modified classical models, where the radiation reaction is deterministic and the power emitted is the classical synchrotron power in the former case and in the latter case accounts for reduction to the power emitted by quantum effects (the Gaunt factor $g$).  We have also considered a stochastic model which calculates the emission using a more physically correct probabilistic treatment.  We have shown that the expectation value of the energy evolves in almost the same way for the stochastic and modified classical models but differently for the classical model.  The variance of the energy distribution evolves differently for all the models.  This suggests that measuring the decrease in the expectation value of the energy is sufficient to measure the Gaunt factor but that a measurement of the variance is required to distinguish quantum stochastic effects.

\ \\
\textbf{Acknowledgements} \ \\
This work was funded by Engineering and Physical Science Research Council grants EP/M018156/1, EP/M018091/1 \& EP/M018555/1 and partially by the Knut and Alice Wallenberg Foundation (TB, MM).  Data access: the data required to reproduce the simulation results presented here is freely available at doi:.  The derivation of the analytical results presented here is given in appendix \ref{moment_derivation}.
\ \\

\appendix

\section{Functions describing synchrotron emission}
\label{special_functions}

The rate of photon emission (making the quasi-static and weak-field approximations) is

\begin{equation}
\label{rate_photons}
\lambda_{\gamma}(\eta) = \frac{\sqrt{3}\alpha_fc}{\lambda_c}\frac{\eta}{\gamma}h(\eta) \quad\quad h(\eta) = \int_0^{\eta/2}d \chi \frac{F(\eta,\chi)}{\chi}. \nonumber
\end{equation}

The quantum synchrotron function is given in \cite{Sokolov_68} eq. (6.5). In our notation it is, for $\chi<\eta/2$,
\begin{equation}
F(\eta,\chi)=\frac{4\chi^2}{\eta^2}y K_{2/3}(y)+
\left(1-\frac{2\chi}{\eta}\right)y\int_y^\infty\,\textrm{d}t\,K_{5/3}(t) \nonumber
\end{equation}

\noindent where $y=4\chi/[3\eta(\eta-2\chi)]$ \& $K_n$ are modified Bessel functions 
of the second kind.  For $\chi\ge \eta/2$, $F(\eta,\chi)=0$. In the classical limit $\hbar\rightarrow0$ the quantum synchrotron spectrum 
reduces to the classical synchrotron spectrum  
$F(\eta,\chi)\rightarrow F_{cl}(y_c)=y_c\int_{y_c}^{\infty}du K_{5/3}(u)$; $y_c=4\chi/3\eta^2$.  The probability that a photon is emitted with a 
given $\chi$ (by an electron with a given $\eta$) is 
$\rho_{\chi}(\eta,\chi)d\chi=[1/h(\eta)][F(\eta,\chi)/\chi]d\chi$.

\section{Derivation of the moment equations}
\label{moment_derivation}

We obtain an equation for the evolution of the expectation value of the electron momentum by multiplying equation (\ref{stochastic_emission_operator}) by $\mathbf{p}$ and integrating over momentum.

\begin{equation}
n_e\left(\frac{d\langle\mathbf{p}\rangle}{dt}\right)_{st}=-\int d^3\mathbf{p}\mathbf{p}\lambda_{\gamma}(\eta)f + \int d^3\mathbf{p} \mathbf{p}\frac{b}{2m_ec}\int_{p}^{\infty}dp'\lambda_{\gamma}(\eta')\rho_{\chi}(\eta',\chi)\frac{p'^2}{p^2}f(\mathbf{p'}). \nonumber
\end{equation}  

\noindent In spherical polars $d^3\mathbf{p} = p^2dpd^2\Omega$.  We also write $\mathbf{p}=p\hat{\mathbf{p}}$. Therefore,

\begin{equation}
n_e\left(\frac{d\langle\mathbf{p}\rangle}{dt}\right)_{st}=-\int d^3\mathbf{p}\mathbf{p}\lambda_{\gamma}(\eta)f + \int d^2\Omega \frac{b\hat{\mathbf{p}}}{2m_ec}\int_0^{\infty}dpp\int_{p}^{\infty}dp'\lambda_{\gamma}(\eta')\rho_{\chi}(\eta',\chi)p'^2f(\mathbf{p'}). \nonumber
\end{equation}

\noindent We may exchange the order of integration over $p$ and $p'$ in the second term on the right-hand side

\begin{equation}
n_e\left(\frac{d\langle\mathbf{p}\rangle}{dt}\right)_{st}=-\int d^3\mathbf{p}\mathbf{p}\lambda_{\gamma}(\eta)f + \int d^2\Omega \frac{b\hat{\mathbf{p}}}{2m_ec}\int_0^{\infty}dp'\lambda_{\gamma}(\eta')f(\mathbf{p'})p'^2 \int_0^{p'}dp p \rho_{\chi}(\eta',\chi). \nonumber
\end{equation}

\noindent Here the $p$ dependence of $\rho_{\chi}$ is in $\chi=[(p'-p)b]/(2m_ec)$ (where we have assumed the electrons are ultra-relativistic).  To simplify the identification of $g P_{cl}$ we define $\rho_{h\nu}dh\nu$ as the probability that an electron with energy parameterised by $\eta$ emits a photon with energy $h\nu$.  $\rho_{\chi}=\rho_{h\nu}(dh\nu/d\chi)=\rho_{h\nu}(2mc^2)/b$. We may therefore write

\begin{equation}
n_e\left(\frac{d\langle\mathbf{p}\rangle}{dt}\right)_{st}=-\int d^3\mathbf{p}\mathbf{p}\lambda_{\gamma}(\eta)f + \int d^2\Omega \hat{\mathbf{p}}\int_0^{\infty}dp'\lambda_{\gamma}(\eta')f(\mathbf{p'})p'^2 \int_0^{p'c}dh\nu \left(p'-\frac{h\nu}{c}\right)\rho_{h\nu}(\eta',h\nu). \nonumber
\end{equation}

\noindent Now we use

\begin{equation}
  \int_0^{p'c}dh\nu \rho_{h\nu}(\eta',h\nu)=1 \quad\quad \int_0^{p'c}dh\nu \rho_{h\nu}(\eta',h\nu)h\nu=(h\nu)_{av} \nonumber
\end{equation}

\noindent to get

\begin{equation}
n_e\left(\frac{d\langle\mathbf{p}\rangle}{dt}\right)_{st}=-\int d^3\mathbf{p}\mathbf{p}\lambda_{\gamma}(\eta)f + \int d^3\mathbf{p}\hat{\mathbf{p}}\lambda_{\gamma}(\eta)f(\mathbf{p})\left(p-\frac{(h\nu)_{av}}{c}\right). \nonumber
\end{equation}

\noindent Cancelling the appropriate terms and identifying $g P_{cl}=\lambda_{\gamma}(h\nu)_{av}$ yields equation (\ref{first_moment_stochastic}),

\begin{equation}
\left(\frac{d\langle\mathbf{p}\rangle}{dt}\right)_{st} = - \frac{\langle gP_{cl}\hat{\mathbf{p}}\rangle}{c}. \nonumber
\end{equation}  

The equation for the evolution of $\sigma^2$ (\ref{second_moment_stochastic}) is obtained by using the same procedure to obtain an equation for $(d \langle \gamma^2\rangle/dt)_{st}$, i.e. we multiply equation (\ref{stochastic_emission_operator}) by $\gamma^2$ and integrate over momentum, 

\begin{equation}
n_e\left(\frac{d\langle\gamma^2\rangle}{dt}\right)_{st}=-\int d^3\mathbf{p}\gamma^2\lambda_{\gamma}(\eta)f + \int d^3\mathbf{p} \gamma^2\frac{b}{2m_ec}\int_{p}^{\infty}dp'\lambda_{\gamma}(\eta')\rho_{\chi}(\eta',\chi)\frac{p'^2}{p^2}f(\mathbf{p'}). \nonumber
\end{equation}  

\noindent Which can be written as

\begin{equation}
n_e\left(\frac{d\langle\gamma^2\rangle}{dt}\right)_{st}=-\int d^3\mathbf{p}\gamma^2\lambda_{\gamma}(\eta)f + \int d^2\Omega\int_0^{\infty}dp'\lambda_{\gamma}(\eta')f(\mathbf{p'})p'^2 \int_0^{p'c}dh\nu \left(\gamma'-\frac{h\nu}{m_ec^2}\right)^2\rho_{h\nu}(\eta',h\nu). \nonumber
\end{equation}

\noindent where we have assumed $\gamma'=p'/m_ec$.  Defining

\begin{equation}
\int_0^{p'c}dh\nu \rho_{h\nu}(\eta',h\nu)(h\nu)^2=[(h\nu)^2]_{av} \nonumber
\end{equation}

\noindent gives

\begin{equation}
n_e\left(\frac{d\langle\gamma^2\rangle}{dt}\right)_{st}=-\int d^3\mathbf{p}\gamma^2\lambda_{\gamma}(\eta)f + \int d^3\mathbf{p}\lambda_{\gamma}(\eta)f(\mathbf{p})\left(\gamma^2-2\gamma\frac{(h\nu)_{av}}{m_ec^2}+\frac{[(h\nu)^2]_{av}}{m_e^2c^4}\right). \nonumber
\end{equation}

\noindent We again cancel the appropriate terms and this time identify $S=\lambda_{\gamma}[(h\nu)^2]_{av}$ as well as $g P_{cl}=\lambda_{\gamma}(h\nu)_{av}$ to get 

\begin{equation}
\left(\frac{d\langle\gamma^2\rangle}{dt}\right)_{st} = - 2\frac{\langle\gamma gP_{cl}\rangle}{m_ec^2} + \frac{\langle S\rangle}{m_e^2c^4}. \nonumber
\end{equation}  

\noindent To get an equation for $(d\sigma^2/dt)_{st}$ we identify $\sigma^2=\langle\gamma^2\rangle - \langle\gamma\rangle^2$.  Therefore,

\begin{equation}
\left(\frac{d\sigma^2}{dt}\right)_{st} = \left(\frac{d\langle\gamma^2\rangle}{dt}\right)_{st} - \left(\frac{d\langle\gamma\rangle^2}{dt}\right)_{st} = \left(\frac{d\langle\gamma^2\rangle}{dt}\right)_{st} - 2\langle\gamma\rangle\left(\frac{d\langle\gamma\rangle}{dt}\right)_{st}. \nonumber
\end{equation}

\noindent Substituting the results for $(d\langle\gamma^2\rangle/dt)_{st}$ and $(d\langle\gamma\rangle/dt)_{st}=\langle g P_{cl}\rangle/(m_ec^2)$  (the latter is obtained by taking the dot product of equation (\ref{first_moment_stochastic}) with $\hat{\mathbf{p}}$ and assuming $p=\gamma m_ec$) gives the result in equation (\ref{second_moment_stochastic}):

\begin{equation}
\left(\frac{d\sigma^2}{dt}\right)_{st} = -2\frac{\langle\gamma gP_{cl}\rangle}{m_ec^2} + \frac{\langle S\rangle}{m_e^2c^4} + 2\langle\gamma\rangle\frac{\langle g P_{cl} \rangle}{m_ec^2} = -2\frac{\langle\Delta\gamma gP_{cl}\rangle}{m_ec^2} + \frac{\langle S\rangle}{m_e^2c^4}. \nonumber
\end{equation}

\noindent Here we have used $\Delta\gamma = \gamma - \langle\gamma\rangle$.

The moments of the classical and modified classical emission operators are straightforwardly obtained by integration by parts.  To obtain equation (\ref{first_moment_deterministic}) for $(d\langle\mathbf{p}\rangle/dt)_{mod \  cl}$ we multiply the emission operator $(\partial f/\partial t)^{mod \ cl}_{em}$ in equation (\ref{deterministic_emission_operators}) by $\mathbf{p}$ and integrate over momentum

\begin{equation}
 n_e\left(\frac{d\langle\mathbf{p}\rangle}{dt}\right)_{mod\ cl} = \int d^3\mathbf{p}\frac{\mathbf{p}}{p^2}\frac{\partial}{\partial p}\left( p^2g \frac{P_{cl}}{c}f \right). \nonumber
\end{equation}

\noindent Substituting $d^3\mathbf{p} = p^2dpd^2\Omega$ and $\mathbf{p} = p\hat{\mathbf{p}}$ and integrating by parts yields

\begin{equation}
 n_e\left(\frac{d\langle\mathbf{p}\rangle}{dt}\right)_{mod\ cl} = \int d^2\Omega \hat{\mathbf{p}}\left(\left[p^3g\frac{P_{cl}}{c}f\right]_0^{\infty}-\int_0^{\infty}dp  p^2g\frac{P_{cl}}{c}f \right) =  - \int d^2\Omega \hat{\mathbf{p}}\int_0^{\infty}dp  p^2g\frac{P_{cl}}{c}f. \nonumber
\end{equation}

\noindent We have used the fact that $f\rightarrow0$ as $p\rightarrow \infty$ (faster than $p^5$ diverges) to get the last result.  We have now derived equation (\ref{first_moment_deterministic})

\begin{equation}
 \left(\frac{d\langle\mathbf{p}\rangle}{dt}\right)_{mod\ cl} =  - \frac{1}{n_e}\int d^3\mathbf{p} g\frac{P_{cl}}{c}\hat{\mathbf{p}}f = -\frac{\langle g P_{cl}\hat{\mathbf{p}}\rangle}{c}. \nonumber
\end{equation}

To derive equation (\ref{second_moment_deterministic}) for $(d\sigma^2/dt)_{mod \  cl}$ we first multiply the emission operator $(\partial f/\partial t)^{mod \ cl}_{em}$ in equation (\ref{deterministic_emission_operators}) by $\gamma^2$ and integrate over momentum

\begin{equation}
 n_e\left(\frac{d\langle\gamma^2\rangle}{dt}\right)_{mod\ cl} = \int d^3\mathbf{p}\frac{\gamma^2}{p^2}\frac{\partial}{\partial p}\left( p^2g \frac{P_{cl}}{c}f \right). \nonumber
\end{equation}

\noindent Substituting $d^3\mathbf{p} = p^2dpd^2\Omega$, $\gamma=p/(m_ec)$ and integrating by parts yields

\begin{equation}
 n_e\left(\frac{d \langle\gamma^2\rangle}{dt}\right)_{mod\ cl} = \int d^2\Omega \left(\left[\frac{p^4}{m_e^2c^2}g\frac{P_{cl}}{c}f\right]_0^{\infty}-2\int_0^{\infty}dp  \frac{p^3}{m_e^2c^2}g\frac{P_{cl}}{c}f \right) =  - \int d^2\Omega \hat{\mathbf{p}}\int_0^{\infty}dp  p^2 \gamma g\frac{P_{cl}}{m_ec^2}f. \nonumber
\end{equation}

\noindent Again, we have used the fact that $f\rightarrow0$ as $p\rightarrow \infty$ (this time faster than $p^6$ diverges) to get the final result.  We may write this more compactly as

\begin{equation}
 \left(\frac{d \langle \gamma^2 \rangle}{dt}\right)_{mod\ cl} =  - \frac{2}{n_e}\int d^3\mathbf{p} \gamma g\frac{P_{cl}}{m_ec^2}\hat{\mathbf{p}}f = -2\frac{\langle\gamma g P_{cl}\rangle}{m_ec^2}. \nonumber
\end{equation}

\noindent We get equation (\ref{second_moment_deterministic}) by identifying $\sigma^2=\langle\gamma^2\rangle - \langle\gamma\rangle^2$ and $\Delta\gamma=\gamma-\langle\gamma\rangle$,

\begin{equation}
\left(\frac{d\sigma^2}{dt}\right)_{mod \ cl} = -2\frac{\langle\gamma gP_{cl}\rangle}{m_ec^2} + 2\langle\gamma\rangle\frac{\langle g P_{cl} \rangle}{m_ec^2} = -2\frac{\langle\Delta\gamma gP_{cl}\rangle}{m_ec^2}. \nonumber
\end{equation}

\section{Derivation of $\xi$}
\label{xi_deriv}

For simplicity in what follows we define $\tau_S$ and $\tau_R$ as

\begin{equation}
S = \frac{m_e^2c^4}{\tau_S}\gamma^4 \quad\quad P_{cl} = \frac{m_ec^2}{\tau_R}\gamma^2. \nonumber
\end{equation} 

\noindent Then we may write $\xi$ as

\begin{equation}
\label{xi_equation}
\xi = \frac{\tau_R}{2\tau_S}\frac{\langle\gamma^4\rangle}{\langle\Delta\gamma\gamma^2\rangle}.
\end{equation}

\noindent where we have set $g_2=g=1$. We may evaluate the averages by substituting $f=[1/(2W\gamma_0m_ec)]\delta(p_y)\delta(p_z)$ for $\gamma_0m_ec(1-W)<p_x<\gamma_0m_ec(1+W)$.

\begin{equation}
\langle\gamma^4\rangle = \frac{1}{2W\gamma_0m_ec}\int_{\gamma_0m_ec(1-W)}^{\gamma_0m_ec(1+W)} \gamma^4dp_x = \frac{\gamma_0^4}{10W}[(1+W)^5-(1-W)^5]=\frac{\gamma_0^4}{5W}(10W^3+5W+W^5)\nonumber
\end{equation} 

\noindent and

\begin{equation}
\langle\Delta\gamma\gamma^2\rangle = \frac{1}{2W\gamma_0m_ec}\int_{\gamma_0m_ec(1-W)}^{\gamma_0m_ec(1+W)} (\gamma-\gamma_0)\gamma^2dp_x = \frac{\gamma_0^3}{24W}[(1-W)^3(1+3W)-(1+W)^3(1-3W)]=\frac{2\gamma_0^3}{3}W^2. \nonumber
\end{equation} 

\noindent Substituting these results into equation (\ref{xi_equation}) yields equation (\ref{T_Q_over_T_C})

\begin{equation}
\xi = \frac{33}{64\sqrt{3}}(10+5W^{-2}+W^2)\eta_0 \approx (3.0+1.5W^{-2}+0.3W^2)\eta_0 \nonumber
\end{equation}

\noindent where we have used $\tau_S/\tau_R = (55b)/(16\sqrt{3})$ and $\eta_0 = \gamma_0 b$.


\bibliography{stochastic_arxiv}

\begin{thebibliography}{}

\bibitem[Andersen et~al., 2012]{Andersen_12}
Andersen, K.~K., Esberg, J., Knudsen, H., Thomsen, H.~D., Uggerh\o{}j, U.~I.,
  Sona, P., Mangiarotti, A., Ketel, T.~J., Dizdar, A., and Ballestrero, S.
  (2012).
\newblock Experimental investigations of synchrotron radiation at the onset of
  the quantum regime.
\newblock {\em Phys. Rev. D.}, 86:072001.

\bibitem[Arber et~al., 2015]{Arber_15}
Arber, T.~D., Bennett, K., Brady, C.~S., Lawrence-Douglas, A., Ramsay, M.~G.,
  Sircombe, N.~J., Gillies, P., Evans, R.~G., Schmitz, H., Bell, A.~R., and
  Ridgers, C.~P. (2015).
\newblock Contemporary particle-in-cell approach to laser-plasma modelling.
\newblock {\em Plasma Phys. and Control. Fusion}, 57:113001.

\bibitem[Baier et~al., 1991]{Baier_98}
Baier, V.~N., Katkov, V.~M., and Strakhovenko, V.~M. (1991).
\newblock Quasiclassical theory of radiation and pair creation in crystals at
  high energy.
\newblock {\em Rad. Eff.}, 527:122--123.

\bibitem[Bashinov and Kim, 2013]{Bashinov_13}
Bashinov, A.~V. and Kim, A.~V. (2013).
\newblock On the electrodynamic model of ultra-relativistic laser-plasma
  interactions caused by radiation reaction effects.
\newblock {\em Physics of Plasmas}, 20(11):113111.

\bibitem[Ba\u{\i}er and Katkov, 1968]{Baier_68}
Ba\u{\i}er, V.~N. and Katkov, V.~M. (1968).
\newblock Quasiclassical theory of bremsstrahlung by relativistic particles.
\newblock {\em Sov. Phys. JETP}, 26:807--813.

\bibitem[Bell and Kirk, 2008]{Bell_08}
Bell, A.~R. and Kirk, J.~G. (2008).
\newblock Possibility of prolific pair production with high-power lasers.
\newblock {\em Phys. Rev. Lett.}, 101:200403.

\bibitem[Blackburn, 2015]{Blackburn_15}
Blackburn, T.~G. (2015).
\newblock Measuring quantum radiation reaction in laser–electron-beam
  collisions.
\newblock {\em Plasma Physics and Controlled Fusion}, 57(7):075012.

\bibitem[Blackburn et~al., 2014]{Blackburn_14}
Blackburn, T.~G., Ridgers, C.~P., Kirk, J.~G., and Bell, A.~R. (2014).
\newblock Quantum radiation reaction in laser\char21{}electron-beam collisions.
\newblock {\em Phys. Rev. Lett.}, 112:015001.

\bibitem[Brady et~al., 2012]{Brady_12}
Brady, C.~S., Ridgers, C.~P., Arber, T.~D., Bell, A.~R., and Kirk, J.~G.
  (2012).
\newblock Laser absorption in relativistically underdense plasmas by
  synchrotron radiation.
\newblock {\em Phys. Rev. Lett.}, 109:245006.

\bibitem[Bula et~al., 1996]{Bula_96}
Bula, C., McDonald, K.~T., Prebys, E.~J., Bamber, C., Boege, S., Kotseroglou,
  T., Melissinos, A.~C., Meyerhofer, D.~D., Ragg, W., Burke, D.~L., Field,
  R.~C., Horton-Smith, G., Odian, A.~C., Spencer, J.~E., Walz, D., Berridge,
  S.~C., Bugg, W.~M., Shmakov, K., and Weidemann, A.~W. (1996).
\newblock Observation of nonlinear effects in compton scattering.
\newblock {\em Phys. Rev. Lett.}, 76:3116--3119.

\bibitem[Bulanov et~al., 2012]{Bulanov_12}
Bulanov, S.~S., Chen, M., Schroeder, C.~B., Esarey, E., Leemans, W.~P.,
  Bulanov, S.~V., Esirkepov, T.~Z., Kando, M., Koga, J.~K., Zhidkov, A.~G.,
  Chen, P., Mur, V.~D., Narozhny, N.~B., Popov, V.~S., Thomas, A. G.~R., and
  Korn, G. (2012).
\newblock On the design of experiments to study extreme field limits
  high-quality electron beams from a laser wakefield accelerator using
  plasma-channel guiding.
\newblock {\em AIP Conf. Proc.}, 1507:825.

\bibitem[Burke et~al., 1997]{Burke_97}
Burke, D.~L., Field, R.~C., Horton-Smith, G., Spencer, J.~E., Walz, D.,
  Berridge, S.~C., Bugg, W.~M., Shmakov, K., Weidemann, A.~W., Bula, C.,
  McDonald, K.~T., Prebys, E.~J., Bamber, C., Boege, S.~J., Koffas, T.,
  Kotseroglou, T., Melissinos, A.~C., Meyerhofer, D.~D., Reis, D.~A., and Ragg,
  W. (1997).
\newblock Positron production in multiphoton light-by-light scattering.
\newblock {\em Phys. Rev. Lett.}, 79:1626--1629.

\bibitem[Di~Piazza et~al., 2010]{DiPiazza_10}
Di~Piazza, A., Hatsagortsyan, K.~Z., and Keitel, C.~H. (2010).
\newblock Quantum radiation reaction effects in multiphoton compton scattering.
\newblock {\em Phys. Rev. Lett.}, 105:220403.

\bibitem[Dinu et~al., 2016]{Dinu_16}
Dinu, V., Harvey, C., Ilderton, A., Marklund, M., and Torgrimsson, G. (2016).
\newblock Quantum radiation reaction: From interference to incoherence.
\newblock {\em Phys. Rev. Lett.}, 116:044801.

\bibitem[DiPiazza, 2008]{DiPiazza_08}
DiPiazza, A. (2008).
\newblock Exact solution of the landau-lifshitz equation in a plane wave.
\newblock {\em Lett. Math. Phys.}, 83:305.

\bibitem[Duclous et~al., 2011]{Duclous_11}
Duclous, R., Kirk, J.~G., and Bell, A.~R. (2011).
\newblock Energy straggling and radiation reaction for magnetic bremsstrahlung.
\newblock {\em Plasma Phys. Control. Fusion}, 53:015009.

\bibitem[Elkina et~al., 2011]{Elkina_11}
Elkina, N.~V., Fedotov, A.~M., Kostyukov, I.~Y., Legkov, M.~V., Narozhny,
  N.~B., Nerush, E.~N., and Ruhl, H. (2011).
\newblock Qed cascades induced by circularly polarized laser fields.
\newblock {\em Phys. Rev. ST Accel. Beams}, 14:054401.

\bibitem[Erber, 1966]{Erber_66}
Erber, T. (1966).
\newblock High-energy electromagnetic conversion processes in intense magnetic
  fields.
\newblock {\em Rev. Mod. Phys.}, 38:626--659.

\bibitem[Faure et~al., 2004]{Faure_04}
Faure, J., Glinec, Y., Pukhov, A., Kiselev, S., Gordienko, S., Lefebvre, E.,
  Rousseau1, J.~P., Burgy, F., and Malka, V. (2004).
\newblock A laser–plasma accelerator producing monoenergetic electron beams.
\newblock {\em Nature}, 431:541--544.

\bibitem[Furry, 1951]{Furry_51}
Furry, W.~H. (1951).
\newblock On bound states and scattering in positron theory.
\newblock {\em Phys. Rev.}, 81:115--124.

\bibitem[Geddes et~al., 2004]{Geddes_04}
Geddes, C. G.~R., Toth, C., van Tilborg, J., Esarey, E., Schroeder, C.~B.,
  Bruhwiler, D., Nieter, C., Cary, J., and Leemans, W.~P. (2004).
\newblock High-quality electron beams from a laser wakefield accelerator using
  plasma-channel guiding.
\newblock {\em Nature}, 431:538--541.

\bibitem[Gonoskov et~al., 2015]{Gonoskov_15}
Gonoskov, A., Bastrakov, S., Efimenko, E., Ilderton, A., Marklund, M., Meyerov,
  I., Muraviev, A., Sergeev, A., Surmin, I., and Wallin, E. (2015).
\newblock Extended particle-in-cell schemes for physics in ultrastrong laser
  fields: Review and developments.
\newblock {\em Phys. Rev. E}, 92:023305.

\bibitem[Harvey et~al., 2017]{Harvey_17}
Harvey, C.~N., Gonoskov, A., Ilderton, A., and Marklund, M. (2017).
\newblock Quantum quenching of radiation losses in short laser pulses.
\newblock {\em Phys. Rev. Lett.}, 118:105004.

\bibitem[Heisenberg and Euler, 1936]{Heisenberg_36}
Heisenberg, W. and Euler, H. (1936).
\newblock Consequences of dirac theory of the positron.
\newblock {\em Z. Phys.}, 98:714.

\bibitem[Ilderton and Torgrimsson, 2013]{Ilderton_13}
Ilderton, A. and Torgrimsson, G. (2013).
\newblock Radiation reaction in strong-field qed.
\newblock {\em Phys. Lett. B}, 725:481.

\bibitem[Kirk et~al., 2009]{Kirk_09}
Kirk, J.~G., Bell, A.~R., and Arka, I. (2009).
\newblock Pair production in counter-propagating laser beams.
\newblock {\em Plasma Phys. Control. Fusion}, 52:085008.

\bibitem[Krivitskii and Tsytovich, 1991]{Krivitskii_91}
Krivitskii, V. and Tsytovich, V. (1991).
\newblock Average radiation-reaction force in quantum electrodynamics.
\newblock {\em Soviet Physics - Uspekhi}, 34(3):250--258.
\newblock cited By 28.

\bibitem[Landau and Lifshitz, 1987]{Landau_87}
Landau, L.~D. and Lifshitz, E.~M. (1987).
\newblock {\em The Classical Theory of Fields}, volume~2 of {\em The Course of
  Theoretical Physics}.
\newblock Butterworth-Heinemann, Oxford.

\bibitem[Leemans et~al., 2014]{Leemans_14}
Leemans, W.~P., Gonsalves, A.~J., Mao, H.-S., Nakamura, K., Benedetti, C.,
  Schroeder, C.~B., T\'oth, C., Daniels, J., Mittelberger, D.~E., Bulanov,
  S.~S., Vay, J.-L., Geddes, C. G.~R., and Esarey, E. (2014).
\newblock Multi-gev electron beams from capillary-discharge-guided subpetawatt
  laser pulses in the self-trapping regime.
\newblock {\em Phys. Rev. Lett.}, 113:245002.

\bibitem[Mangles et~al., 2004]{Mangles_04}
Mangles, S., Murphy, C.~D., Najmudin, Z., Thomas, A. G.~R., Collier, J.~L.,
  Dangor, A.~E., Divall, E.~J., Foster, P.~S., Gallacher, J.~G., Hooker, C.~J.,
  Jaroszynski, D.~A., Langley, A.~J., Mori, W.~B., Norreys, P.~A., Tsung,
  F.~S., Viskup, R., Walton, B.~R., and Krushelnick, K. (2004).
\newblock Monoenergetic beams of relativistic electrons from intense
  laser–plasma interactions.
\newblock {\em Nature}, 431:535--538.

\bibitem[Neitz and Di~Piazza, 2013]{Neitz_13}
Neitz, N. and Di~Piazza, A. (2013).
\newblock Stochasticity effects in quantum radiation reaction.
\newblock {\em Phys. Rev. Lett.}, 111:054802.

\bibitem[Ridgers et~al., 2014]{Ridgers_14}
Ridgers, C.~P., Kirk, J.~G., Duclous, R., Blackburn, T.~G., Brady, C.~S.,
  Bennett, K., Arber, T.~D., and Bell, A.~R. (2014).
\newblock Pair production in counter-propagating laser beams.
\newblock {\em J. Comput. Phys.}, 260:273.

\bibitem[Ritus, 1985]{Ritus_85}
Ritus, V.~I. (1985).
\newblock Quantum effects of the interaction of elementary particles with an
  intense electromagnetic field.
\newblock {\em J. Russ. Laser Res.}, 6:497.

\bibitem[Sarri et~al., 2014]{Sarri_14}
Sarri, G., Corvan, D.~J., Schumaker, W., Cole, J.~M., Di~Piazza, A., Ahmed, H.,
  Harvey, C., Keitel, C.~H., Krushelnick, K., Mangles, S. P.~D., Najmudin, Z.,
  Symes, D., Thomas, A. G.~R., Yeung, M., Zhao, Z., and Zepf, M. (2014).
\newblock Ultrahigh brilliance multi-mev $\ensuremath{\gamma}$-ray beams from
  nonlinear relativistic thomson scattering.
\newblock {\em Phys. Rev. Lett.}, 113:224801.

\bibitem[Shen and White, 1972]{Shen_72}
Shen, C.~S. and White, D. (1972).
\newblock Energy straggling and radiation reaction for magnetic bremsstrahlung.
\newblock {\em Phys. Rev. Lett.}, 28:455.

\bibitem[Sokolov and Ternov, 1968]{Sokolov_68}
Sokolov, A.~A. and Ternov, I.~M. (1968).
\newblock {\em Synchrotron Radiation}.
\newblock Akademie-Verlag, Berlin.

\bibitem[Sokolov et~al., 2010]{Sokolov_10_2}
Sokolov, I.~V., Naumova, N.~M., Nees, J.~A., and Mourou, G.~A. (2010).
\newblock Pair creation in qed-strong pulsed laser fields interacting with
  electron beams.
\newblock {\em Phys. Rev. Lett.}, 105:195005.

\bibitem[Tajima and Dawson, 1979]{Tajima_79}
Tajima, T. and Dawson, J.~M. (1979).
\newblock Laser electron accelerator.
\newblock {\em Phys. Rev. Lett.}, 43:267--270.

\bibitem[Thomas et~al., 2012]{Thomas_12}
Thomas, A. G.~R., Ridgers, C.~P., Bulanov, S.~S., Griffin, B.~J., and Mangles,
  S. P.~D. (2012).
\newblock Strong radiation-damping effects in a gamma-ray source generated by
  the interaction of a high-intensity laser with a wakefield-accelerated
  electron beam.
\newblock {\em Phys. Rev. X}, 2:041004.

\bibitem[Vranic et~al., 2015]{Vranic_15}
Vranic, M., Grismayer, T., Fonseca, R.~A., and Silva, L.~O. (2015).
\newblock Quantum radiation reaction in head-on laser-electron beam
  interaction.
\newblock {\em New J. Phys.}, 18:073035.

\bibitem[Vranic et~al., 2014]{Vranic_14}
Vranic, M., Martins, J.~L., Vieira, J., Fonseca, R.~A., and Silva, L.~O.
  (2014).
\newblock All-optical radiation reaction at $1{0}^{21}\text{ }\text{
  }\mathrm{W}/{\mathrm{cm}}^{2}$.
\newblock {\em Phys. Rev. Lett.}, 113:134801.

\bibitem[Wang et~al., 2015]{Wang_15}
Wang, H.~Y., Yan, X.~Q., and Zepf, M. (2015).
\newblock Signatures of quantum radiation reaction in laser-electron-beam
  collisions.
\newblock {\em Physics of Plasmas}, 22(9):093103.

\bibitem[Zhang et~al., 2015]{Zhang_15}
Zhang, P., Ridgers, C.~P., and Thomas, A. G.~R. (2015).
\newblock The effect of nonlinear quantum electrodynamics on relativistic
  transparency and laser absorption in ultra-relativistic plasmas.
\newblock {\em New J. Phys.}, 27:043051.

\end{thebibliography}


\bibliographystyle{apalike}


\end{document}